\documentclass[12pt]{article}

\begin{document}

\title{Saturation of Electrostatic Potential: Exactly Solvable 
2D Coulomb Models}

\author{L. {\v S}amaj$^1$}

\maketitle

\begin{abstract}
We test the concepts of renormalized charge and potential saturation, 
introduced within the framework of highly asymmetric Coulomb mixtures, 
on exactly solvable Coulomb models.
The object of study is the average electrostatic potential induced by 
a unique ``guest'' charge immersed in a classical electrolyte, the whole 
system being in thermal equilibrium at some inverse temperature $\beta$.
The guest charge is considered to be either an infinite hard wall carrying 
a uniform surface charge or a charged colloidal particle.
The systems are treated as two-dimensional; the electrolyte is modelled 
by a symmetric two-component plasma of point-like $\pm e$ charges 
with logarithmic Coulomb interactions.
Two cases are solved exactly: the Debye-H\"uckel limit 
$\beta e^2\to 0$ and the Thirring free-fermion point $\beta e^2 = 2$.
The results at the free-fermion point can be summarized as follows:
(i) The induced electrostatic potential exhibits the asymptotic behavior, 
at large distances from the guest charge, whose form is different from that
obtained in the Debye-H\"uckel (linear Poisson-Boltzmann) theory.
This means that the concept of renormalized charge, developed within the 
nonlinear Poisson-Boltzmann theory to describe the screening effect
of the electrolyte cloud, fails at the free-fermion point.  
(ii) In the limit of an infinite bare charge, the induced electrostatic 
potential saturates at a finite value in every point of the electrolyte region.
This fact confirms the previously proposed hypothesis of potential saturation.
\end{abstract}

\medskip

\noindent {\bf KEY WORDS:} Coulomb systems; colloids; charge renormalization;
electrostatic potential saturation; solvable models.

\vfill

\noindent $^1$ 
Institute of Physics, Slovak Academy of Sciences,
D\'ubravsk\'a cesta 9, 845 11 Bratislava, Slovak Republic;
e-mail: fyzimaes@savba.sk

\newpage

\renewcommand{\theequation}{1.\arabic{equation}}
\setcounter{equation}{0}

\section{Introduction}
Asymmetric classical Coulomb mixtures, such as highly charged colloidal or 
polyelectrolyte suspensions, in the strong coupling regime, 
have attracted much attention in the last years 
(for a review of phenomenological approaches built on the base 
of mean-field theories, see ref. \cite{Levin}).
The concept of renormalized charge has been introduced within the 
Wigner-Seitz cell models to describe an effective interaction between 
highly-charged ``macro-ions'' as a result of their strong positional 
correlations with the oppositely charged ``micro-ions'' 
\cite{Manning,Alexander,Lowen,Trizac1,Trizac2}.

The concept of renormalized charge can be documented in the infinite
dilution limit of colloids \cite{Belloni,Diehl,Trizac3,Bocquet,Auboy}.
The simplified model consists of a unique colloidal particle idealized
as a hard sphere of radius $a$ carrying charge $Z e$, $Z$ is the valence 
and $e$ the elementary charge, at its centre (any spherically symmetric
charge distribution inside the colloid can be represented in this way).
The colloid is immersed in an electrolyte modelled by a symmetric 
two-component plasma (TCP) of elementary $\pm e$
(if possible, point-like) charges. 
The system is defined in an infinite $\nu$-dimensional space of points
${\bf r}\in R^{\nu}$, having for simplicity vacuum dielectric constant
$\epsilon = 1$.
The interaction energy of two particles of charges $q$ and $q'$ at
the respective spatial positions ${\bf r}$ and ${\bf r}'$ is given by
$ q q' \phi(\vert {\bf r}-{\bf r}'\vert)$, where $\phi$, the Coulomb 
potential induced by a unit charge, is the solution of the Poisson equation
\begin{equation} \label{1.1}
\Delta \phi({\bf r}) = - s_{\nu} \delta({\bf r})
\end{equation} 
$s_{\nu}$ is the surface area of the $\nu$-dimensional unit sphere.
This definition of the $\nu$-dimensional Coulomb potential maintains 
many generic properties (e.g., screening sum rules \cite{Martin}) 
of ``real'' three-dimensional (3D) Coulomb systems 
with the interaction potential $\phi(r) = 1/r, r\in R^3$.
In particular, in 2D,
\begin{equation} \label{1.2}
\phi({\bf r}) = - {\rm ln}(\vert {\bf r}\vert/r_0),
\qquad {\bf r}\in R^2
\end{equation}
where $r_0$ is a free length scale.
Thermal equilibrium is treated in the grand canonical ensemble
characterized by the inverse temperature $\beta=1/(kT)$ and by the
couple of bulk particle fugacities for electrolyte $\pm e$-charged 
particles, $z_+ = z_- = z$.
The corresponding average bulk densities will be denoted by 
$n_+ = n_- = n/2$, $n$ is the total number density.
For the classical case of point-like particles, the singularity of the Coulomb
potential $\phi({\bf r})$ at the origin ${\bf r}={\bf 0}$ often prevents
the thermodynamic stability against the collapse of positive-negative
pairs of charges: in 2D, for small enough temperatures; in 3D,
for any finite temperature.

We shall explain the concept of renormalized charge on the 3D mean-field 
theories, valid in the high-temperature limit and free from the
collapse problem of point-like charges.
Let us fix the colloidal particle at the origin ${\bf 0}$ and denote by 
$\psi({\bf r})$ the induced average electrostatic potential at point ${\bf r}$.
Inside the colloidal hard-core region $0< r \le a$, $\psi({\bf r})$
satisfies the Poisson equation
\begin{equation} \label{1.3}
\Delta \psi_{<}({\bf r}) = - 4 \pi Z e \delta({\bf r})
\end{equation}
Inside the electrolyte region $r > a$, the Poisson equation takes the form
\begin{equation} \label{1.4}
\Delta \psi_{>}({\bf r}) = - 4 \pi \rho({\bf r})
\end{equation}
where
\begin{equation} \label{1.5}
\rho({\bf r}) = e \left[ n_+({\bf r}) - n_-({\bf r}) \right]
\end{equation}
is the charge density of electrolyte particles.
Within the ordinary mean-field approach, the average particle densities
at a given point are approximated by replacing the potential of 
mean force by the average electrostatic potential at that point,
$n_{\pm}({\bf r}) = n_{\pm} \exp[ \mp \beta e \psi_>({\bf r})]$.
Eq. (\ref{1.4}) then reduces to the non-linear Poisson-Boltzmann (PB)
equation
\begin{equation} \label{1.6}
\Delta \psi_>({\bf r}) = 4 \pi e n \, 
{\rm sinh}\left[ \beta e \psi_>({\bf r}) \right]
\end{equation}
Debye and H\"uckel proposed a linearization of this equation by considering 
the small-argument expansion of 
${\rm sinh}(\beta e \psi_>)\sim \beta e \psi_>$  
valid in the high-temperature limit.
The linear PB equation then reads
\begin{equation} \label{1.7}
\left( \Delta - \kappa^2 \right) \psi_>({\bf r}) = 0
\end{equation}
where $\kappa=\sqrt{4\pi\beta e^2 n}$ is the 3D inverse Debye length
of the electrolyte particles.
Due to the spherical symmetry of the problem,
$\Delta = \partial_r^2 + (2/r) \partial_r$.
Eqs. (\ref{1.3}) and (\ref{1.7}), subject to the condition of
regularity at $r\to\infty$ and the boundary conditions of continuity
of the electrostatic potential and the electric field across the
colloid surface $r=a$, imply that, in the linear Debye-H\"uckel theory,
\begin{eqnarray}
\psi_<(r) & = & Ze \left( \frac{1}{r} - \frac{\kappa}{1+\kappa a}
\right) \label{1.8} \\
\psi_>(r) & = & \frac{Z e}{1+\kappa a} 
\frac{\exp\left[-\kappa(r-a)\right]}{r} \label{1.9}
\end{eqnarray}
The charge density of electrolyte particles is given by
\begin{equation} \label{1.10}
\rho(r) = - \beta e^2 n \psi_>(r), \qquad r>a
\end{equation}
Strictly speaking, the linearization of Eq. (\ref{1.6}) is only valid
if $\beta e \psi({\bf r})<<1$, and this is indeed true for asymptotically
large distances $r$ where the screened potential $\psi({\bf r})$ vanishes.
As a consequence, at large $r$, the solution of the non-linear PB
equation (\ref{1.6}) must also take the Yukawa form
\begin{equation} \label{1.11}
\psi_>(r) \sim A \frac{\exp\left[-\kappa(r-a)\right]}{r},
\qquad r\to\infty
\end{equation}
with $A$ being an $r$-independent constant.
The non-linearity of the problem is reflected only through an effective 
boundary condition at the colloid surface determining the constant $A$.
Comparing (\ref{1.11}) with (\ref{1.9}) one sees that $A$ is related to
the renormalized colloid valence, $Z_{\rm ren}$, as follows
\begin{equation} \label{1.12}
A = \frac{Z_{\rm ren} e}{1+\kappa a}
\end{equation}
The renormalized charge reflects the screening effect of the electrolyte 
cloud, and can further be used to establish an effective interaction for 
a system of guest charges immersed in the electrolyte.
Since the exact solution of the non-linear PB equation (\ref{1.6})
for the sphere geometry is not available, the constant $A$ can be 
determined only approximately, i.e., in the limiting case $\kappa a >> 1$, 
by matching with the exact solution of the non-linear PB for 
the charged-plane geometry \cite{Trizac3,Bocquet}.
An important feature is that, as expected from the Manning condensation
theory \cite{Manning}, the renormalized valence saturates at some finite 
value $Z_{\rm ren}^{\rm sat}$ when the colloidal bare valence $Z$ 
goes to infinity.

More refined approaches, which go beyond the mean-field approximation
and incorporate electrostatic correlations among the electrolyte
particles, were developed in refs. \cite{Groot,Barbosa}. 
Monte-Carlo simulations \cite{Groot} indicate the existence of
a maximum in the plot of the renormalized charge versus the bare
colloidal charge.

As was correctly mentioned by T\'ellez and Trizac \cite{Tellez},
the definition of renormalized charge requires that the average
electrostatic potential behaves far from the colloid as it would
within the linearized Debye-H\"uckel theory, up to the constant
prefactor.
This is not at all ensured for a finite temperature.
To avoid this artificial limitation in the saturation problem, one considers 
the possibility of a more general phenomenon of {\em potential} saturation: 
the question is whether, in the limit of the infinite bare colloidal charge 
$Z e\to\infty$, the induced average electrostatic potential 
$\psi^{\rm sat}({\bf r})$ is finite inside the whole electrolyte region $r>a$
\footnote{based on simple electrostatics, the potential is infinite 
at the colloid surface $r=a$}.
We would like to emphasize that the potential saturation, if it exists,
is the pure non-linearity effect: there is no potential saturation within 
the linear Debye-H\"uckel theory, see Eq. (\ref{1.9}).

It is well-known that the linearized Debye-H\"uckel theory correctly
describes the small-coupling (high-temperature) limit $\beta e^2\to 0$
in the sense that the basic screening properties of the charged system
in the conducting regime are preserved.
In the present case one can readily verify by using Eqs. (\ref{1.9})
and (\ref{1.10}) that the screening cloud of the electrolyte particles 
compensates exactly the bare charge of the ``guest'' colloid:
\begin{equation} \label{1.13}
Z e + \int_a^{\infty} {\rm d}r\, 4\pi r^2 \rho(r) = 0
\end{equation}
This sum rule no longer holds within the non-linear PB theory.
This serious deficiency of the non-linear PB theory is ``tolerated''
in various phenomenological approaches \cite{Levin} because of
the predicted charge renormalization which is in a relatively
good agreement with available experimental data.

The present work aims to put the concept of charge renormalization and 
the hypothesis of the electric potential saturation on a rigorous basis.
As a test model for the electrolyte, we use the 2D symmetric TCP of
point-like $\pm e$ charges with logarithmic pairwise interactions (\ref{1.2}).
The 2D plasma of point-like charges is stable against the collapse 
of positive-negative pairs of charges provided that the corresponding 
Boltzmann factor $r^{-\beta e^2}$ is integrable at short distances in 2D, 
i.e. for the (dimensionless) coupling constant $\beta e^2 < 2$.
In this stability range of couplings, the equilibrium statistical mechanics
of the plasma (the bulk thermodynamics, special cases of the surface
thermodynamics and the large-distance behavior of the two-body correlation
function) is exactly solvable via an equivalence with the integrable 2D 
Euclidean sine-Gordon field theory (for a short review, see ref. \cite{Samaj}).
The complete exact information about correlation functions is available in two 
special cases: in the high-temperature Debye-H\"uckel limit $\beta e^2\to 0$,
and just at the collapse point $\beta e^2 = 2$ \cite{Cornu1,Cornu2} 
which corresponds to the free-fermion point of an equivalent 2D Thirring 
model (although the free energy and the particle density diverge at the
collapse point, the truncated Ursell correlation functions are finite).
We examine the above-outlined problems in these two exactly solvable cases,
for two particular geometries: the charged line and the charged circular 
colloid.
Based on the exact results we show that the concept of renormalized charge
does not apply to the studied 2D microscopic Coulomb system.
On the other hand, the anticipated phenomenon of electric potential
saturation is confirmed.

The paper is organized as follows.
Section 2 deals with the charged-line geometry.
Section 2.1 is devoted to a short recapitulation of the Debye-H\"uckel
limit.
In Section 2.2, the known exact results at the free-fermion point 
\cite{Cornu2} are analyzed from the point of view of the studied subjects.
Section 3 deals with the colloidal-charge geometry.
As before, Section 3.1 concerns the Debye-H\"uckel limit, while
Section 3.2 is devoted to an original derivation of the exact solution
at the free-fermion point.
A recapitulation and some concluding remarks are given in Section 4. 

\renewcommand{\theequation}{2.\arabic{equation}}
\setcounter{equation}{0}

\section{Charged line}
We consider an infinite 2D space of points ${\bf r}\in R^2$ defined by
Cartesian coordinates $(x,y)$.
The half-space $x<0$, impenetrable to particles, is assumed to be 
a vacuum hard wall.
The electrolyte of $\pm e$ point-like charges is confined to the
complementary half-space $x>0$.
The model interface is the line localized at $x=0$, along the $y$-axis.
The line, which carries a uniform charge $\sigma e$ per unit length,
models an electrode.
There is another electrode of opposite charge density localized at
$x = +\infty$.
The electrostatic potential induced by the two electrodes is $0$ for
$x<0$ and $-2\pi \sigma e x$ for $x>0$.
The boundary condition for the electric field reads
\begin{equation} \label{2.1}
- \frac{{\rm d} \psi(x)}{{\rm d} x}\Bigg\vert_{x=0} = 2\pi \sigma e
\end{equation}

\subsection{Debye-H\"uckel limit}
The average electrostatic potential at distance $x$ form the interface
satisfies the 2D Poisson equation
\begin{equation} \label{2.2}
\frac{{\rm d}^2 \psi(x)}{{\rm d}x^2} = - 2 \pi \rho(x) ,
\qquad x\ge 0
\end{equation}
In the spirit of the Debye-H\"uckel theory valid in the limit 
$\beta e^2 \to 0$, the charge density of the electrolyte particles 
is approximated, in close analogy with Eq. (\ref{1.10}), 
as $\rho(x) \sim -\beta e^2 n \psi(x)$.
The linear PB equation thus reads
\begin{equation} \label{2.3}
\left( \frac{{\rm d}^2}{{\rm d}x^2} - \kappa^2 \right) \psi(x)
= 0 , \qquad x\ge 0
\end{equation}
where $\kappa = \sqrt{2\pi\beta e^2 n}$ is the 2D inverse Debye length.
The solution of (\ref{2.3}), subject to the requirement of regularity
at $x\to\infty$ and the boundary condition (\ref{2.1}), takes the form
\begin{equation} \label{2.4}
\psi(x) = \frac{2\pi \sigma e}{\kappa} \exp(-\kappa x) ,
\qquad x\ge 0
\end{equation}
The consequent charge density
\begin{equation} \label{2.5}
\rho(x) = - \sigma e \kappa \exp(-\kappa x) , \qquad x\ge 0
\end{equation}
fulfills the following analogue of the screening sum rule (\ref{1.13}):
\begin{equation} \label{2.6}
\sigma e + \int_0^{\infty} {\rm d}x\, \rho(x) = 0
\end{equation}

\subsection{The free-fermion point}
Since the particle density diverges at the collapse point $\beta e^2 = 2$,
the available thermodynamic parameter is the fugacity $z$.
It will be considered in a rescaled form, $m = 2\pi r_0 z$ [$r_0$ is the
length scale considered in (\ref{1.2})], having dimension of an inverse length.
The density profiles of electrolyte particles near the charged plain hard 
wall were obtained in ref. \cite{Cornu2} by solving the Green-function 
problem of the corresponding boundary Thirring model at its free-fermion point.
The result is
\begin{eqnarray} 
n_{\pm}(x) & = & n_{\pm} + \frac{m^2}{4\pi} 
\left[ K_2(2 m x) - K_0(2 m x) \right] \nonumber \\
& & - \frac{m^2}{2\pi} \left[ \frac{1}{2 m x} + \frac{1}{(2 m x)^2} \right] 
\exp(- 2 m x) \label{2.7} \\
& & + \frac{m^2}{2\pi} \int_0^{\mp 2\pi\sigma} 
\frac{{\rm d}t}{\sqrt{m^2+t^2}-t} \exp\left( - 2 \sqrt{m^2+t^2} x \right)
\nonumber
\end{eqnarray}
where $K_l$ are the modified Bessel functions of order $l$.
The divergent bulk particle densities $n_+ = n_- = n/2$ can be regularized,
e.g., by considering a small hard core around each particle \cite{Cornu2};
since we are interested in the charge density $\rho(x)$ defined by 
the difference $e[n_+(x) - n_-(x)]$, we avoid this regularization procedure.
After some simple algebra, one gets
\begin{equation} \label{2.8}
\rho(x) = - \frac{e}{\pi} \int_0^{2\pi\sigma} {\rm d}t\, 
\sqrt{m^2+t^2} \exp\left( - 2 \sqrt{m^2+t^2} x \right)
\end{equation}
It is easy to check that the screening sum rule (\ref{2.6}) is fulfilled
by this charge density.

The corresponding electrostatic potential, determined by the Poisson equation 
(\ref{2.2}) and the requirement of regularity at $x\to\infty$, reads
\begin{equation} \label{2.9}
\psi(x) = \frac{e}{2} \int_0^{2\pi\sigma} \frac{{\rm d}t}{\sqrt{m^2+t^2}} 
\exp\left( - 2 \sqrt{m^2+t^2} x \right)
\end{equation}
The boundary condition (\ref{2.1}) is evidently satisfied for this potential.
In order to obtain the large-$x$ expansion of $\psi(x)$, we first make 
in (\ref{2.9}) a change of the integration variable $t$ into 
$u = x[\sqrt{1+(t/m)^2}-1]$, and then expand 
the integrated function in powers of $1/x$.
In the leading order,
\begin{equation} \label{2.10}
\psi(x) \sim \frac{e}{4} \left( \frac{\pi}{m x} \right)^{1/2}
\exp(-2 m x) , \qquad x\to\infty
\end{equation}
for all $\sigma\ne 0$ 
[$\sigma=0$ implies the trivial result $\psi(x)\equiv 0$].
The independence of the leading asymptotic term (\ref{2.10}) on the (nonzero) 
charge density $\sigma$ is a special feature of the present geometry.
Comparing (\ref{2.10}) to the Debye-H\"uckel result (\ref{2.4}) 
characterized by the pure exponential decay in $x$, and identifying 
the respective inverse lengths $2 m$ and $\kappa$, one sees that 
the large-$x$ behaviors differ one from the other.
The idea of renormalized charge density thus fails.
On the other side, considering the limit of the dimensionless ratio
$\sigma/m\to\infty$ in (\ref{2.9}), one has explicitly
\begin{equation} \label{2.11}
\psi^{\rm sat}(x) = \frac{e}{2} K_0( 2 m x )
\end{equation}
It follows from the basic properties of $K_0$ that 
$0\le \psi^{\rm sat}(x) < \infty$ for all $x>0$, in full agreement
with the saturation hypothesis.

\renewcommand{\theequation}{3.\arabic{equation}}
\setcounter{equation}{0}

\section{Charged circular colloid}
As in the 3D case discussed in the Introduction, we fix at the origin 
one colloid of charge $Z e$ and disk hard core with radius $a$.
There is an infinite 2D TCP of $\pm e$ point-like charges in 
the complementary outer space. 
The analogue of the boundary condition (\ref{2.1}) for the
electric field now reads
\begin{equation} \label{3.1}
- \frac{\partial \psi({\bf r})}{\partial r}\Bigg\vert_{r=a}
= \frac{Z e}{a}
\end{equation}

\subsection{Debye-H\"uckel limit}
Inside the colloidal hard-core region $0<r\le a$, the electrostatic
potential $\psi({\bf r})$ satisfies the 2D Poisson equation
\begin{equation} \label{3.2}
\Delta \psi_<({\bf r}) = - 2\pi Z e \delta({\bf r})
\end{equation}
Inside the electrolyte region $r>a$, the consideration of the linear 
mean-field prescription $\rho({\bf r}) \sim - \beta e^2 n \psi({\bf r})$ 
in the Poisson equation implies
\begin{equation} \label{3.3}
\left( \Delta - \kappa^2 \right) \psi_>({\bf r}) = 0
\end{equation}
Due to the circular symmetry of the problem,
$\Delta = \partial_r^2 + (1/r) \partial_r$.
Eqs. (\ref{3.2}) and (\ref{3.3}), subject to the requirement of
regularity at $r\to\infty$ and the usual boundary conditions of
continuity across the colloid boundary $r=a$, imply
\begin{eqnarray}
\psi_<(r) & = & Z e \left[ - {\rm ln} \left( \frac{r}{a} \right)
+ \frac{K_0(\kappa a)}{\kappa a K_1(\kappa a)} \right] \label{3.4} \\
\psi_>(r) & = & \frac{Z e}{\kappa a K_1(\kappa a)} K_0(\kappa r)
\label{3.5} 
\end{eqnarray}
The boundary condition (\ref{3.1}) is trivially satisfied.
Note that, after defining the surface charge density $\sigma e = Z e/(2\pi a)$
and going to the limits $a,r\to\infty$ with a fixed difference
$r-a = x>0$, (\ref{3.5}) reduces to the straight-line result (\ref{2.4})
as it should be.
At large $r$, using the asymptotic formula for $K_0$ \cite{Gradshteyn},
the average electrostatic potential (\ref{3.5}) behaves like
\begin{equation} \label{3.6}
\psi_>(r) \sim \frac{Z e}{\kappa a K_1(\kappa a)}
\left( \frac{\pi}{2\kappa r} \right)^{1/2}
\exp(-\kappa r), \qquad r\to\infty
\end{equation}
The electrolyte charge density, given by
\begin{equation} \label{3.7}
\rho(r) = - \frac{Z e \kappa}{2\pi a K_1(\kappa a)} K_0(\kappa r),
\qquad r>a
\end{equation}
fulfills the screening sum rule
\begin{equation} \label{3.8}
Z e + \int_a^{\infty} {\rm d}r\, 2\pi r \rho(r) = 0
\end{equation}

\subsection{The free-fermion point}
According to the general formalism established in ref. \cite{Cornu1,Cornu2},
in order to obtain density profiles of electrolyte $\pm e$ particles
at coupling $\beta e^2 = 2$, one has to solve the Green function problem
of a $2\times 2$ matrix ${\bf G}({\bf r},{\bf r}')$.
Its matrix elements $G_{qq'}({\bf r},{\bf r}')$ ($q,q'=\pm$ denote the charge 
sign) are determined by a system of four coupled partial differential 
equations (PDE), written in a $2\times 2$ matrix notation as follows
\begin{equation} \label{3.9}
\left[ \sigma^1 \partial_x + \sigma^2 \partial_y + 
m_+({\bf r}) \frac{{\bf 1}+\sigma^3}{2} + 
m_-({\bf r}) \frac{{\bf 1}-\sigma^3}{2} \right]
{\bf G}({\bf r},{\bf r}') = {\bf 1} \delta({\bf r}-{\bf r}') 
\end{equation} 
Here, ${\bf 1}$ and $\sigma^i$ $(i=1,2,3)$ denote the $2\times 2$ unit
and Pauli matrices, respectively, and
\begin{equation} \label{3.10}
m_q({\bf r}) = m({\bf r}) \exp\left[ - 2 q v({\bf r}) \right],
\qquad q=\pm
\end{equation}
is the position-dependent (rescaled) fugacity for some external
electric potential $v({\bf r})$ (in units of $e$); 
a nonelectric potential which acts in the same way on both kinds 
of particles, like an impenetrable hard wall or core, 
is described by the region-dependence of $m({\bf r})$.
Four Eqs. (\ref{3.9}) split into two independent sets of equations,
the one for the pair $(G_{++},G_{-+})$ and the other for the pair
$(G_{--},G_{+-})$.
We shall present a detailed derivation of the results for the pair
$(G_{++},G_{-+})$, given by
\begin{eqnarray}
m_+({\bf r}) G_{++}({\bf r},{\bf r}') + 
\left( \partial_x - {\rm i} \partial_y \right) G_{-+}({\bf r},{\bf r}')
& = & \delta({\bf r}-{\bf r}') \label{3.11} \\
\left( \partial_x + {\rm i} \partial_y \right) G_{++}({\bf r},{\bf r}')
+ m_-({\bf r}) G_{-+}({\bf r},{\bf r}') & = & 0 \label{3.12}
\end{eqnarray} 
Based on a similar derivation procedure, we shall only present the final 
results for the pair $(G_{--},G_{+-})$.
As concerns the boundary conditions, since Eq. (\ref{3.9}) is a first-order
system, all matrix elements $G_{qq'}({\bf r},{\bf r}')$ must be continuous
when crossing a boundary between two different regions.
The requirement of regularity is obvious.
The one-particle densities are given by
\begin{equation} \label{3.13}
n_q({\bf r}) = m_q({\bf r}) \lim_{{\bf r}'\to {\bf r}}
G_{qq}({\bf r},{\bf r}') , \qquad q = \pm
\end{equation}
In the bulk regime with $m({\bf r}) = m$ for all points ${\bf r}\in R^2$,
one has \cite{Cornu2}
\begin{equation} \label{3.14}
G_{qq}({\bf r},{\bf r}')= \frac{m}{2\pi} K_0(m\vert {\bf r}-{\bf r}' \vert) ,
\qquad q=\pm
\end{equation}
so the one-particle densities diverge logarithmically as 
${\bf r}'\to {\bf r}$ in (\ref{3.13}).
This divergence can be suppressed by introducing a short-distance
cut-off $R$, 
\begin{equation} \label{3.15}
n_{\pm} = \lim_{m R\to 0} \frac{m^2}{2\pi} K_0( mR )
\sim \frac{m^2}{\pi} \left[ {\rm ln} \left( \frac{2}{m R} \right) - C \right]
\end{equation}
where $C$ is Euler's constant.

In the case of the studied model, the colloidal particle at the origin 
induces the electrostatic potential (in units of $e$) 
$v({\bf r}) = - Z {\rm ln}(r/r_0)$.
The rescaled fugacity $m({\bf r})=0$ inside the hard-core region $0<r\le a$ 
and $m({\bf r}) = m$ in the electrolyte region $r>a$.
Thus,
\begin{equation} \label{3.16}
m_{\pm}({\bf r}) =  \cases{0&for $0<r\le a$\cr
& \cr
m (r/r_0)^{\pm 2 Z}&for $r>a$}
\end{equation}
For our purpose it will be sufficient to consider the source point ${\bf r}'$ 
in Eqs. (\ref{3.11}) and (\ref{3.12}) to be localized only in 
the electrolyte region, so, without writing it explicitly, in what follows 
we shall assume that $r'>a$.
As concerns the point ${\bf r}$, let us first assume its localization
in the colloidal hard-core region, i.e. $r\le a$.
Taking $m_{\pm}({\bf r}) = 0$ in Eqs. (\ref{3.11}) and (\ref{3.12}), one gets
\begin{eqnarray}
\left( \partial_x + {\rm i} \partial_y \right) 
G_{++}({\bf r},{\bf r}') & = & 0, \qquad r\le a \label{3.17} \\
\left( \partial_x - {\rm i} \partial_y \right) 
G_{-+}({\bf r},{\bf r}') & = & 0, \qquad r\le a \label{3.18} 
\end{eqnarray}
This means that, as functions of ${\bf r}$, $G_{++}$ depends only on 
$z = r \exp({\rm i}\varphi)$ and $G_{-+}$ depends only on
${\bar z} = r \exp(-{\rm i}\varphi)$.  
The general solutions of Eqs. (\ref{3.17}) and (\ref{3.18}), regular at $r=0$, 
can be therefore written as the expansions of the forms
\begin{eqnarray}
G_{++}({\bf r},{\bf r}') & = & \frac{m}{2\pi} \sum_{l=0}^{\infty} 
f_l(m r',\varphi') (m r)^l \exp({\rm i}l\varphi) , \qquad r\le a 
\label{3.19} \\
G_{-+}({\bf r},{\bf r}') & = & \frac{m}{2\pi} \sum_{l=0}^{\infty} 
h_l(m r',\varphi') (m r)^l \exp( -{\rm i}l\varphi ) , \qquad r\le a 
\label{3.20}
\end{eqnarray}
where the functions $\{ f_l, h_l \}$ are determined by the boundary
conditions at $r=a$.
When the point ${\bf r}$ is localized in the electrolyte region, i.e. $r>a$, 
taking $m_{\pm}({\bf r}) = m (r/r_0)^{\pm 2Z}$, we first express by using 
Eq. (\ref{3.12}) $G_{-+}$ in terms of $G_{++}$:
\begin{equation} \label{3.21}
G_{-+}({\bf r},{\bf r}') = - \frac{1}{m} \left( \frac{r}{r_0} \right)^{2Z}
\left( \partial_x + {\rm i} \partial_y \right) G_{++}({\bf r},{\bf r}'),
\qquad r>a
\end{equation}
The consequent substitution of $G_{-+}$ into (\ref{3.11}) implies
the only PDE determining $G_{++}$.
After lengthy but simple algebra, in terms of the auxiliary two-point function
\begin{equation} \label{3.22}
g_{++}({\bf r},{\bf r}') = - \frac{1}{m} \left( \frac{r}{r_0} \right)^Z
G_{++}({\bf r},{\bf r}') \left( \frac{r'}{r_0} \right)^Z 
\end{equation}
this PDE is obtained in the form
\begin{equation} \label{3.23}
\left( - m^2 - {\hat H} \right) g_{++}({\bf r},{\bf r}')
= \delta({\bf r}-{\bf r}'), \qquad r>a
\end{equation}
where
\begin{equation} \label{3.24}
{\hat H} = - \Delta_{\bf r} - \frac{2 Z {\rm i}}{r^2} 
\left( x \partial_y - y \partial_x \right) + \frac{Z^2}{r^2}
\end{equation}
It is clear that $g_{++}({\bf r},{\bf r}') = 
\langle {\bf r} \vert (-m^2-{\hat H})^{-1} \vert {\bf r}' \rangle$
is nothing but the Green-function two-point matrix element associated
with the one-particle quantum Hamiltonian ${\hat H}$ and the spectral
parameter $-m^2$.
In polar coordinates $(r,\varphi)$, the Hamiltonian (\ref{3.24}) reads
\begin{equation} \label{3.25}
{\hat H} = - \frac{\partial^2}{\partial r^2} - 
\frac{1}{r}\frac{\partial}{\partial r} -
\frac{1}{r^2}\frac{\partial^2}{\partial\varphi^2}
- \frac{2 Z {\rm i}}{r^2} \frac{\partial}{\partial\varphi} + \frac{Z^2}{r^2}
\end{equation}
According to elementary quantum mechanics, the periodicity requirement 
under the shift $\varphi\to \varphi + 2\pi$ implies that the eigenfunctions 
of ${\hat H}$ have a trivial dependence on the angle $\varphi$:
$\Psi_l \propto \exp({\rm i}l\varphi)$ where $l=0,\pm 1,\ldots$ is
the ``magnetic'' quantum number.
It follows from Eq. (\ref{3.25}) that the radial part of the eigenfunction
with a given $l$ is then determined by
\begin{equation} \label{3.26}
{\hat H}_l = - \frac{\partial^2}{\partial r^2} - 
\frac{1}{r} \frac{\partial}{\partial r} + \frac{(l+Z)^2}{r^2}
\end{equation}
This is the radial Hamiltonian of a free quantum particle in 2D,
where the presence of the colloidal charge $Z e$ manifests itself 
as the shift of the quantum number $l$ by the integer $Z$.
The standard Green-function technique implies an explicit form of $g_{++}$.
Using then the relation (\ref{3.22}), $G_{++}({\bf r},{\bf r}')$ 
with $r,r'>a$ is found to be
\begin{eqnarray}
G_{++}({\bf r},{\bf r}') & = & \frac{m}{2\pi}
\left( \frac{r_0}{r} \right)^Z \left( \frac{r_0}{r'} \right)^Z
\sum_{l=-\infty}^{\infty} \exp\left[ {\rm i}l(\varphi-\varphi') \right] 
\nonumber \\ & \times &
[ I_{l+Z}(mr_<) K_{l+Z}(mr_>) + c_l K_{l+Z}(mr) K_{l+Z}(m r') ] 
\label{3.27}
\end{eqnarray} 
Here, $I_l$ and $K_l$ are the modified Bessel functions, and
$r_<$ ($r_>$) is the smaller (the larger) of $r$ and $r'$. 
$G_{-+}$ is generated from $G_{++}$ via Eq. (\ref{3.21}).
For the special case $a<r<r'$, it takes the form
\begin{eqnarray}
G_{-+}({\bf r},{\bf r}') & = & \frac{m}{2\pi}
\left( \frac{r}{r_0} \right)^Z \left( \frac{r_0}{r'} \right)^Z
\sum_{l=-\infty}^{\infty} 
\exp\left[ {\rm i}(l+1)\varphi-{\rm i}l\varphi' \right] 
\nonumber \\ & \times &
[- I_{l+Z+1}(mr) + c_l K_{l+Z+1}(mr)] K_{l+Z}(m r')
\label{3.28}
\end{eqnarray} 
The unknown constants $\{ c_l \}$ are determined by the requirements
that $G_{++}$ and $G_{-+}$ be continuous at $r=a$.
With regard to Eqs. (\ref{3.19}) and (\ref{3.20}), one gets
\begin{equation} \label{3.29}
c_l =  \cases{-I_{l+Z}(ma)/K_{l+Z}(ma)&$l < 0$\cr
& \cr
I_{l+Z+1}(ma)/K_{l+Z+1}(ma)&$l\ge 0$}
\end{equation}
Applying the ``summation theorem'' for the modified Bessel functions
\cite{Gradshteyn}
\begin{equation} \label{3.30}
K_0(m\vert {\bf r}-{\bf r}'\vert) = \sum_{l=-\infty}^{\infty}
\exp[{\rm i}l(\varphi-\varphi')] I_l(m r_<) K_l(m r_>)
\end{equation}
we conclude that, when both points ${\bf r}$ and ${\bf r}'$ are
in the electrolyte region $(r,r'>a)$,
\begin{eqnarray}
G_{++}({\bf r},{\bf r}') & = & \frac{m}{2\pi}
\left( \frac{r_0}{r} \right)^Z \left( \frac{r_0}{r'} \right)^Z
\Bigg\{ \exp\left[- {\rm i}Z(\varphi-\varphi') \right]
K_0(m\vert {\bf r}-{\bf r}'\vert) \nonumber \\
& & + \sum_{l\ge 0} \exp[{\rm i}l(\varphi-\varphi')]
\frac{I_{l+Z+1}(ma)}{K_{l+Z+1}(ma)} K_{l+Z}(mr) K_{l+Z}(mr') 
\phantom{space}\label{3.31} \\  
& & - \sum_{l<0} \exp[{\rm i}l(\varphi-\varphi')]
\frac{I_{l+Z}(ma)}{K_{l+Z}(ma)} K_{l+Z}(mr) K_{l+Z}(mr')\Bigg\} \nonumber
\end{eqnarray}

The matrix element $G_{--}({\bf r},{\bf r}')$ can be deduced in
the same way.
It is expressible by formula (\ref{3.31}) under the substitution $Z\to -Z$, 
which corresponds to the sign reversal of the colloidal charge.

The densities of electrolyte particles at $r>a$ follow from
the relation (\ref{3.13}):
\begin{equation} \label{3.32}
n_{\pm}(r) =  n_{\pm} + \frac{m^2}{2\pi} \sum_{l=\pm Z}^{\infty}
\frac{I_{l+1}(ma)}{K_{l+1}(ma)} [K_l(mr)]^2 
- \frac{m^2}{2\pi} \sum_{l=-\infty}^{\pm Z-1}
\frac{I_l(ma)}{K_l(ma)} [K_l(mr)]^2 
\end{equation}
Using the Wronskian relation \cite{Gradshteyn}
\begin{equation} \label{3.33}
I_l(x) K_{l+1}(x) + I_{l+1}(x) K_l(x) = \frac{1}{x}
\end{equation}
together with the symmetry properties $I_l = I_{-l}$ and $K_l = K_{-l}$
for integer $l$, the charge density in the electrolyte region is found 
in the form
\begin{equation} \label{3.34}
\rho(r) = - \frac{m e}{2\pi a} \sum_{l=1}^Z
\frac{1}{K_{l-1}(ma) K_l(ma)} \left( [K_{l-1}(mr)]^2 + [K_l(mr)]^2 \right),
\qquad r>a
\end{equation}
It stands to reason that $\rho(r)=0$ in the colloidal region $r\le a$
due to the hard-core repulsion.
With the aid of the integral formula
\begin{equation} \label{3.35}
\int_a^{\infty} {\rm d}r\, r [K_l(mr)]^2 = \frac{a^2}{2}
\left( K_{l-1}(ma) K_{l+1}(ma) - [K_l(ma)]^2 \right)
\end{equation}
and the recursion relation \cite{Gradshteyn}
\begin{equation} \label{3.36}
x [K_{l-1}(x) - K_{l+1}(x)] = - 2 l K_l(x)
\end{equation}
taken at $l=Z$, the charge density (\ref{3.34}) can be shown to fulfill
the screening sum rule (\ref{3.8}).

The average electrostatic potential is given by 
the couple of Poisson equations
\begin{eqnarray}
\Delta \psi_<({\bf r}) & = & - 2 \pi Z e \delta({\bf r}),
\qquad r \le a \label{3.37} \\
\Delta \psi_>({\bf r}) & = & - 2 \pi \rho({\bf r}),
\qquad r>a \label{3.38}
\end{eqnarray}
The circularly symmetric solution of Eq. (\ref{3.37}) reads
\begin{equation} \label{3.39}
\psi_<(r) =  Z e \left[ - {\rm ln} \left( \frac{r}{a} \right) + {\rm const}
\right]
\end{equation}
The circularly symmetric potential of Eq. (\ref{3.38}) 
fulfills the differential equation
\begin{equation} \label{3.40}
\frac{{\rm d}}{{\rm d}r} \left[ r \frac{{\rm d}\psi_>(r)}{{\rm d}r} \right]
= \frac{m e}{a} \sum_{l=1}^Z \frac{1}{K_{l-1}(ma) K_l(ma)} 
\, r \left( [K_{l-1}(mr)]^2 + [K_l(mr)]^2 \right)
\end{equation}
The first integration of Eq. (\ref{3.40}) with respect to $r$ 
can be performed easily.
Using the regularity condition $\lim_{r\to\infty} r\, {\rm d}_r \psi_>(r) = 0$,
the integration formula of type (\ref{3.35}) and finally the recursion
relation (\ref{3.36}), one arrives at
\begin{equation} \label{3.41}
\frac{{\rm d}\psi_>(r)}{{\rm d}r} = - \frac{e}{a}
\sum_{l=1}^Z \frac{K_{l-1}(mr) K_l(mr)}{K_{l-1}(ma) K_l(ma)}
\end{equation}
Note the obvious fulfillment of the boundary condition (\ref{3.1}).
The subsequent integration of Eq. (\ref{3.41}) with respect to $r$ 
is a bit more complicated problem.
The regularity condition $\lim_{r\to\infty} \psi_>(r) = 0$ has to be combined 
with the integral formula [derivable by using the relation (\ref{3.52})]
\begin{equation} \label{3.42}
m \int_r^{\infty} {\rm d}r'\, K_{l-1}(mr') K_l(mr')
= (-1)^{l+1} \frac{1}{2} \sum_{j=0}^{l-1} (-1)^j \mu_j
[K_j(mr)]^2, \quad l\ge 1
\end{equation}
$\mu_j$ is the Neumann factor: $\mu_0=1$ and $\mu_j=2$ for $j\ge 1$,
to get
\begin{equation} \label{3.43}
\psi_>(r) = - \frac{e}{2} \sum_{j=0}^{Z-1} (-1)^j \mu_j f_j(ma)
[ K_j(mr) ]^2
\end{equation}
where
\begin{equation} \label{3.44}
f_j(ma) = \sum_{l=j+1}^Z (-1)^l \frac{1}{ma} \frac{1}{K_{l-1}(ma) K_l(ma)}
\end{equation}
With regard to the Wronskian relation (\ref{3.33}), $f_j$ can be
simplified as follows
\begin{eqnarray}
f_j(ma) & = & \sum_{l=j+1}^Z (-1)^l \left[ \frac{I_l(ma)}{K_l(ma)} 
+ \frac{I_{l-1}(ma)}{K_{l-1}(ma)} \right] \nonumber \\
& = & (-1)^Z \frac{I_Z(ma)}{K_Z(ma)} - (-1)^j \frac{I_j(ma)}{K_j(ma)}
\label{3.45}
\end{eqnarray}
Thus, in the electrolyte region $r>a$,
\begin{eqnarray}
\psi_>(r) & = & \frac{e}{2} \sum_{j=0}^{Z-1} \mu_j \frac{I_j(ma)}{K_j(ma)}
[K_j(mr)]^2 \nonumber \\
& & + \frac{e}{2} (-1)^{Z+1} \frac{I_Z(ma)}{K_Z(ma)}
\sum_{j=0}^{Z-1} (-1)^j \mu_j [K_j(mr)]^2  \label{3.46} 
\end{eqnarray}

Before analyzing the result (\ref{3.46}), let us recall some basic properties 
of the modified Bessel functions $I_l(x)$ and $K_l(x)$ $(l=0,1,\ldots)$ 
when the argument $x$ belongs to the interval $0\le x < \infty$.
$I_l(x)$ and $K_l(x)$ satisfy the same differential equation,
\begin{equation} \label{3.47}
\frac{{\rm d}^2f}{{\rm d}x^2} + \frac{1}{x}\frac{{\rm d}f}{{\rm d}x}
- \left( 1 + \frac{l^2}{x^2} \right) f = 0, \qquad
\mbox{$f = I_l(x)$ or $K_l(x)$} 
\end{equation}
but exhibit different asymptotic behaviors:
\begin{equation} \label{3.48}
I_l(x) \sim \frac{1}{\sqrt{2\pi x}}\, {\rm e}^x , \quad
K_l(x) \sim \left( \frac{\pi}{2 x} \right)^{1/2} {\rm e}^{-x}
\qquad \mbox{for $x\to\infty$}
\end{equation}
and 
\begin{equation} \label{3.49}
I_l(x) \sim \frac{1}{l!} \left( \frac{x}{2} \right)^l , \quad
K_l(x) \sim \frac{(l-1)!}{2} \left( \frac{x}{2} \right)^{-l}
\qquad \mbox{for $x\to 0$}
\end{equation}
except for the special $l=0$ case of $K_0(x)\sim -{\rm ln}(x/2)-C$.
Both $I_l(x)$ and $K_l(x)$ are positive for $x\ge 0$.
In particular: $I_l(x)$ starts from 0 at $x=0$ [except for the special 
case of $I_0(0)=1$] and monotonously increases to infinity at $x\to\infty$; 
$K_l(x)$ is infinite at $x=0$ and monotonously decreases to zero 
at $x\to\infty$.

For finite $Z$, at large $r$, the average electrostatic potential
(\ref{3.46}) behaves like
\begin{equation} \label{3.50}
\psi_>(r) \sim \frac{\pi e}{4} \left[ \sum_{j=0}^{Z-1} \mu_j
\frac{I_j(ma)}{K_j(ma)} + \frac{I_Z(ma)}{K_Z(ma)} \right]
\frac{1}{mr} \exp(-2 m r), \qquad r\to\infty
\end{equation}
This asymptotic behavior differs from the large-distance prediction 
(\ref{3.6}) of the Debye-H\"uckel theory by the factor $r^{-1/2}$, 
which is in contradiction with the concept of renormalized charge.

In order to resolve the saturation problem in the limit $Z\to\infty$,
we first use the integral formula (\ref{3.42}) to rewrite the last
term on the rhs of Eq. (\ref{3.46}) as follows
\begin{eqnarray}
\psi_>(r) & = & \frac{e}{2} \sum_{j=0}^{Z-1} \mu_j \frac{I_j(ma)}{K_j(ma)}
[K_j(mr)]^2 \nonumber \\
& & + e \frac{I_Z(ma)}{K_Z(ma)} m \int_r^{\infty} {\rm d}r'\,
K_{Z-1}(m r') K_Z(m r') \label{3.51}
\end{eqnarray}
The positivity of the modified Bessel functions then ensures that,
for every $Z$, $\psi_>(r)\ge 0$ in the whole region $r>a$.
In order to establish an upper bound for $\psi_>(r)$, we first use
the relation \cite{Gradshteyn}
\begin{equation} \label{3.52}
K_{l-1}(x) + K_{l+1}(x) = - 2 \frac{{\rm d} K_l(x)}{{\rm d}x}
\end{equation} 
to establish the equality 
\begin{equation} \label{3.53}
m \int_r^{\infty} {\rm d}r'\, K_{Z-1}(m r') K_Z(m r') = [ K_Z(mr) ]^2 
- m \int_r^{\infty} {\rm d}r'\, K_Z(m r') K_{Z+1}(m r') 
\end{equation}
The consideration of this equality in Eq. (\ref{3.51}) leads to
\begin{equation} \label{3.54}
\psi_>(r) \le \frac{e}{2} \sum_{j=-Z}^Z \frac{I_j(ma)}{K_j(ma)} [ K_j(mr) ]^2 ,
\qquad r > a
\end{equation}
In the considered region $r>a$, the inequality $K_j(mr)/K_j(ma) < 1$
implies that
\begin{equation} \label{3.55}
\psi_>(r) < \frac{e}{2} \sum_{j=-Z}^Z I_j(ma) K_j(mr) , \qquad r>a
\end{equation}
In the limit $Z\to\infty$, the application of the summation theorem
(\ref{3.30}) with $\varphi=\varphi'$ finally gives
\begin{equation} \label{3.56}
0 \le \psi^{\rm sat}(r) < \frac{e}{2} K_0\left( m(r-a) \right) , \qquad r>a
\end{equation}
The existence of the lower and upper bounds for $\psi^{\rm sat}$
in the electrolyte region confirms once again the validity of 
the potential saturation hypothesis \cite{Tellez}.

\section{Conclusion}
We have studied the average electrostatic potential
induced by a unique ``guest'' charge immersed in an infinite electrolyte,
the electrolyte being modelled by the classical two-component plasma
of elementary $\pm e$ point-like charges.
The primary motivation came from the predictions of the two basic
3D mean-field theories described in the Introduction: the Debye-H\"uckel 
theory based on the linear PB equation and the non-linear PB theory.
The important point is that both mean-field theories predict the same type
behavior of the induced electrostatic potential at asymptotically large 
distances from the guest charge, only the constant prefactors are different.
Within the non-linear PB theory, this fact permits one to introduce 
the renormalized guest charge which involves the non-linear screening 
effect of the electrolyte cloud, and can further be used to establish 
an effective interaction for a system of guest charges immersed in
the electrolyte.
When the bare guest charge goes to infinity, the renormalized charge
saturates at a finite value.

In order to go beyond the mean-field methods, we have have tested
the concept of charge renormalization on the 2D Coulomb models.
These system have advantage of being completely solvable not only 
in the Debye-H\"uckel high-temperature limit $\beta e^2\to 0$, 
but also at a finite temperature, namely the Thirring free-fermion
point $\beta e^2 = 2$.
Although just at this point the collapse of positive-negative pairs
of point-like charges emerges, the charge-density profile in 
the electrolyte region (determining the average electrostatic potential
through the Poisson equation) is a well-defined finite quantity
which satisfies the electroneutrality sum rule.
We have considered two geometries of the guest charge:
the infinite hard wall carrying the uniform surface charge (Section 2)
and the charged colloidal particle with a hard core (Section 3).
For both geometries, the results at the free-fermion point 
can be summarized as follows.
The asymptotic large-distance behavior of the induced electrostatic potential 
differs from that predicted by the linear Debye-H\"uckel theory, so 
the concept of renormalized charge does not apply.
On the other hand, when the bare guest charge goes to infinity, the
induced potential saturates at a finite value in each point of the
electrolyte region.
In the case of the infinite charged wall, the saturation potential
was found explicitly, see Eq. (\ref{2.11}). 
In the more complicated case of the charged colloidal particle, 
lower and upper bounds for the saturated potential were derived, 
see formula (\ref{3.56}).
These results confirm that the potential saturation hypothesis
\cite{Tellez} is indeed true.

It is an open question whether the failure of the concept of renormalized
charge is restricted to the free-fermion point, or it takes place also in 
some subinterval of the stability region $0<\beta e^2 <2$.
This question might be answered by exploring a form-factor approach to the 
integrable 2D sine-Gordon model, the field-theory equivalent of the 2D TCP.
Without going into detail, the form-factor method serves as a tool for
a systematic generation of the large-distance asymptotic expansion 
for two-point correlation functions.
Its recent developments seem to be applicable to the present problems.
We shall proceed along this line.

\section*{Acknowledgments}
I thank Bernard Jancovici and Emmanuel Trizac for very useful comments.
The support by Grant VEGA 2/3107/24 is acknowledged.

\end{document}